\documentclass[prl,twocolumn]{revtex4} 
\usepackage{graphicx}
\usepackage{amssymb,amsmath}
\usepackage{xcolor}

\begin{document}

\title{The $c-d$ conjecture}

\author{Jos\'e I. Latorre$^{1,2}$ and Germ\'an  Sierra$^3$ \vspace{0.2cm} \\  
${}^1$ Centre for Quantum Technologies, National University of Singapore, Singapore. \\
${}^2$ Quantum Research Centre, Technology Innovation Institute, Abu Dhabi, United Arab Emirates. \\
 ${}^3$  Instituto de F\'isica Te\'orica UAM-CSIC, Universidad Aut\'onoma de Madrid, Spain.}



\bigskip\bigskip\bigskip\bigskip

\begin{abstract}
We conjecture  a relation between the local dimension $d$ of a local nearest-neighbor 
critical Hamiltonian  in one spatial dimension and the maximum central charge, $c_{\text{max}}$, that it can yield. 
Specifically, we propose that $c_{\text{max}} \leq d-1$, 
 establishing a link between the short-distance lattice realization of a model and its emerging long-distance entanglement properties. 
 This inequality can be viewed as a general form of  a $c$-theorem establishing the reduction of effective degrees of freedom between the 
 UV lattice and the IR conformal field theory.
We support this conjecture with numerous examples. 
\end{abstract}


\maketitle

\vskip 0.2cm

\def\barray{\begin{eqnarray}}
\def\earray{\end{eqnarray}}
\def\beq{\begin{equation}}
\def\eeq{\end{equation}}
\def\cmax{c_{\text{max}}}

\subsection{Introduction}

Quantum systems defined on a lattice can develop long-distance correlations that decay as a power law at zero temperature, depending on the parameters of the Hamiltonian of the system. This phenomenon corresponds to a quantum phase transition which is controlled by the properties of the conformal group. 

The emergence of quantum correlations that decay algebraically from the very detailed structure of local interactions remains a subtle issue, where few solid theorems are available. Indeed, it is not fully understood how local interactions keep extending entanglement in a structured way so that correlators decay smoothly and the entanglement entropy obeys scaling laws. 

Here, we shall present a conjecture that relates the local dimension of a one-dimensional lattice Hamiltonian to the central charge of the conformal theory that emerges at long distances. It is, thus, a relation of ultraviolet (UV) physics with its infrared (IR) realization. 
To be precise, we conjecture that a nearest-neighbor critical Hamiltonian, that is translational invariant, where the local degrees of freedom span $d$ dimensions, can only give rise to a conformal theory central charge limited by
\beq
 \cmax \le d-1 .  
 \label{1} 
\eeq 
That is, no arbitrarily large central charge can be obtained from a nearest-neighbor Hamiltonian of finite local dimension. 
We shall name this proposal as the $c-d$ conjecture. 

Although the $c-d$ conjecture seems physically plausible, its validity isn't immediately evident.
The dimension $d$ is related to the dimensionality of the local Hilbert space and shows up in the original lattice Hamiltonian, which is a dimensionful operator.  On the other side, the central charge $c$ can be determined from various sources. These include extracting it from the ground state via the von Neumann entropy of the reduced density matrix \cite{CW94,H94,V03,CC04,K04}, identifying it through finite size corrections of the ground state energy or free energy \cite{BCN86,A86}, or through correlators associated with the energy momentum tensor \cite{BPZ84}. These options require having solved the system. They appear as an emergent  property of the theory. Alternatively the central charge is a key element in the algebra of operators dictated by the conformal group. The relation between operators in the lattice and conformal operators in the long-distance regime
 is non-trivial \cite{KS94,MV17}. This avenue is definitely relevant to have a complete proof of the $c-d$ conjecture, if true. 

A different way to see the non-trivial nature of the $c-d$ conjecture arises from the existence of local, translational-invariant Hamiltonians that deliver ground states that are maximally entangled, representing superpositions of Motzkin paths \cite{MS16,B12}.
The ground state of the model displays volume law entanglement, showing that this property can be achieved 
through local Hamiltonians of dimension $d \geq 3$. Yet, the model is no longer conformal and escapes the conjecture. The lesson to be taken from this example is that the dimension $d$ has the potential to distribute significantly larger entanglement than that found in a conformal field theory. Any proof of the $c-d$ conjecture should then make explicit use of the properties of conformal field theories.

\subsection{The $c-d$ conjecture}

There is a simple heuristic argument to understand that no arbitrary large central charge can emerge from finite local dimensions. We may argue this fact using the scaling of entanglement entropy. It is known that the ground state of a critical Hamiltonian carries a von Neumann entropy for a 
reduced density matrix to $L$ sites that scales as $S\sim \frac{c}{3} \log L$ \cite{V03,CC04}. On the other hand, if the system is made of a chain of degrees of freedom of dimension $d$, the maximum entropy of a set of $L$ sites is bounded by $S\le L \log d$. Necessarily $S\sim \frac{c}{3} \log L < L \log d$. It is then clear that for any large size, it is impossible to accommodate an arbitrary large central charge with a finite $d$. This argument is yet too coarse to bring further insight into the way that a finite value of $d$ limits the amount of long distance conformal correlations that can be achieved. 


It is important to again emphasize the requirement of conformal symmetry. As mentioned before, local interactions can deliver volume law entanglement, which is far stronger than the area law observed in  conformal field theory. In particular, it is known that for spin $j\ge 3/2$ there are local interactions whose ground state is described by large superpositions of Motskin paths \cite{MS16,B12}. These hamiltonians are not conformal theories and have ground states that display volume law entanglement. The role of conformal invariance is then seen as a complex combination of constraints that have to be provided by $d$ locals degrees of freedom, hence the limitation that $\cmax \leq d-1$.

Let us now formulate the $c-d$ conjecture in more detail.
Let us consider a one-dimensional quantum lattice system, with a local
Hilbert space of dimension $d$, that is ${\cal H}_1 =  \mathbb{C}^{\otimes d}$, and $L > 2$ sites. 
The total Hilbert space is  ${\cal H}_{L}  = {\cal H}_1^{ \otimes L}$. 
We define a Hamiltonian acting on ${\cal H}_{L}$
\barray 
H_L & = &  \sum_{i=1}^L \, h_{i, i+1}, \qquad h_{L,L+1} =  h_{L,1} \, ,  \label{3}   \\
h_{i, i+1} & = & {\bf 1} \otimes \stackrel{ i-1}{ \dots} \otimes \;   h  \; \otimes  \stackrel{ L- i-1}{ \dots} \otimes {\bf 1},
\quad i =1, \dots, L-1 \, , 
\nonumber 
\earray 
where $h$ is an hermitean $d^2 \times d^2$ matrix. 
We have assumed periodic boundary conditions,  expressed as $h_{L,L+1} = h_{L,1}$,
ensuring  that $H_L$ remains unchanged under the translation of the lattice by one site.

\vspace{0.5 cm}

{\bf Conjecture}: If the  low energy spectrum of the Hamiltonian \eqref{3} is described
by a collection of unitary CFTs with central charges $c_a$, then
\beq
c_T   \leq d -1,\qquad  {\rm where} \quad c_T = \sum_a  c_a\  .
\label{4}
\eeq

The case $d=1$ is evidently fulfilled since  there is only a single state in ${\cal H}_L$
in a unitary CFT with  $c=0$ that corresponds to the vacuum.   Another consistency check of Eq.(\ref{4}) 
is to consider the tensor product of two quantum  models  with local dimensions $d_1$ and $d_2$,
each of them satisfying Eq.(\ref{4}), that is  $\sum_a  c_{1,a} \leq d_1 -1$ and
$\sum_b  c_{2,b} \leq d_2 -1$. Compounding the two theories as a direct product  gives 
\beq
\sum_a  c_{1,a}  + \sum_b c_{2,b}  \leq d_1 + d_2 -2 <
d_1  d_2 -1,
\eeq
since the direct product Hilbert space has dimension $d_1 d_2$.

Let us note here the apparent similarity of the $c-d$ conjecture in Eq.(\ref{4}) with the
$c$-theorem \cite{Zam86}. The statement of irreversibility of renormalization group flows \cite{Zam86} states that in two space-time dimensions 
the UV fixed point carries a central charge $c_{UV}$ which is an upper bound for the one in the IR $c_{IR}$, that is $c_{IR} \leq c_{UV}$. The $c$-theorem is formulated for 
relativistic QFTs, so it cannot be straightforwardly  applied to our problem. 
We may however interpret Eq.(\ref{4}) as an expanded version of the $c$-theorem where
$c_{UV}$ is replaced by $d-1$.  An additional difficulty to  use QFT techniques in trying to prove
Eq.(\ref{4}) is that the  low energy physics  of the Hamiltonian (\ref{3}) may correspond to
CFTs with  different speeds of propagation, say  $v_a$, 
as in the case of the Hubbard model, away from half filling,  where the spin wave velocity $v_s$ 
and charge velocity $v_c$ can be different \cite{Hubbard}.

The $c-d$ conjecture is, in its present form, limited to nearest-neighbor interactions. 
It is easy to see that non nearest-neighbor interactions can be handled as nearest-neighbor ones on a larger local Hilbert space. 
As an example, consider a Hamiltonian with next-to-nearest-neighbor interactions $h_{i,i+1,i+2}$.
 Assuming an even number of sites $L$, we can group the sites into pairs $(2i-1,2i)$ and define a nearest-neighbor Hamiltonian for 
$L/2$ sites, given by  
\beq
h_{(2i-1,2i),  (2i+1, 2i+2)} = h_{2i-1,2i, 2i+1}+ h_{2i,2i+1,2i+2}
\eeq
with  local dimension is $d^2$. 
 In this context, the $c-d$ conjecture leads to the inequality   $c \leq d^2-1$.

Let us make the observation that the $c-d$ conjecture is limited by some bounds. This can be argued 
by analyzing how the combination of two theories works. Let us accept that $c_{max} = f(d)$. Two independent theories can
be compounded onto a single theory of larger dimension. That is, if $c_{max,1}=f(d_1)$ and $c_{max,2}=f(d_2)$,
it follows that 
\beq
   f(d_1) + f(d_2) \leq f(d_1\times d_2)\ ,
   \label{convex}
\eeq
\noindent 
as the dimension of the total Hilbert space is the product of the dimensions of the two separate ones. Let us further assume that  the asymptotic scaling of $f(d)$ is a power law with exponent $\alpha$. Form Eq.(\ref{convex}), it follows that
\beq
f(d)\sim d^\alpha\longrightarrow \alpha \geq \log 2 / \log d. 
\label{scaling}
\eeq
For $d=2$, the case of spin 1/2 chains, the bound becomes $\alpha\geq 1$. This bound becomes weaker as the dimension $d$ grows. As a matter of fact, 
in the limit of very large $d$, the functionality $f(d)=\log d$
saturates Eq.(\ref{scaling}). We shall show later on that some families of models do  follow  this log relation.

\subsection{Evidence supporting the $c-d$ conjecture}

The $c-d$ conjecture can be tested on a large series of known models. 
Some non-trivial examples of Eq.(\ref{4}) are shown in the following table, which strongly suggests that it must hold in general. 

\[ 
\begin{array}{|c|c|c|c|c|} 
\hline
d & c & {\rm Model}   & {\rm CFT} & {\rm Sym} \\ 
\hline
\hline
2& \frac{1}{2}  & {\rm Ising} & {\cal M}_{3,4} & \mathbb{Z}_2  \\ 
\hline
3& \frac{4}{5}  & 3 \;  {\rm state} \; {\rm Potts}  & {\cal M}_{5,6} & Z_3   \\ 
\hline 
4 & 1 & 4  \; {\rm state} \; {\rm Potts}  & {\cal M}_{\infty,\infty} & Z_4 \\ 
\hline  
2 & 1  &   XXZ_{1/2}     & U(1) \;    & U(1)    \\ 
\hline 
4& 1 &   {\rm Ashkin -  Teller}   & U(1) \;  {\rm  orbifold}   & \mathbb{Z}_2 \otimes \mathbb{Z}_2     \\ 
\hline  
Q & \frac{2(Q-1)}{Q+2}   &   {\rm Parafermion} \,     &  {\rm Parafermion}  & \mathbb{Z}_Q   \\ 
\hline 
3 & \frac{3}{2} ,   1  &   {\rm FZ}    & U(1) \;    & U(1)    \\ 
\hline 
2 s +1& 
\frac{3 s}{s+1} 
&   XXX_s   & SU(2)_{2s} \;    & SU(2)     \\ 
\hline
3 & 
\frac{1}{2} ,   \frac{7}{10} &   {\rm Blume -  Capel}    & \; {\cal M}_{3,4}, {\cal M}_{4,5}    & \mathbb{Z}_2   \\ 
\hline 
3 & 
\begin{array}{c}
3/2 \\
2 \\
\end{array}  & \begin{array}{c}
{\rm spin }\, 1 \,  {\rm bilinear}  \\
- {\rm biquadratic} \\
\end{array}  & \begin{array}{c}
SU(2)_2 \\
SU(3)_1   \\
\end{array} & SO(3)  \\
\hline 
N & N-1  & SU(N) &  SU(N)_1 & SU(N)  \\
\hline 
\end{array}
\]

Table: Quantum lattice models with nearest-neighbour interations with local dimension $d$, central charge $c$,
associated CFT and symmetries of the Hamiltonian.  ${\cal M}_{m,m+1}$ denotes
the minimal model with central charge $c = 1- \frac{6}{m (m+1)}$, and $G_k$ a Wess-Zumino-Witten model associated to a Lie algebra $G$ and level $k$. 

\vspace{0.4cm} 

The details of each model, including Hamiltonian, central charges and related symmetries are presented in the Supplementary Material section. 

It is interesting to notice that the $c-d$ conjecture is saturated in some cases. For instance, for $d=2$, the spin 1/2 antiferromagnetic Heisenberg model  has central charge $c=1$, thus obeying $c=d-1$. According to the conjecture there is no other model made with spins that can provide a larger central charge. 

The saturation of the $c-d$ conjecture bound is also reached by the critical $SU(N)$ spin chain where 
$d=N$ and $c=N-1$. When  $N=2$ it coincides with the result previously discussed.

\subsection{Discussion}

The $c-d$ conjecture provides a relation between the UV and IR realizations of a theory. The building blocks of a model may display some symmetry among UV degrees of freedom. At long distances, the effective IR realization of the theory may describe some extra symmetry such as conformal invariance. The $c-d$ conjecture proposes a strict bound on the amount of entanglement which is present in the IR conformal realization of a theory.

The $c-d$ conjecture bares some relation to other efforts to relate UV with IR realizations of a theory. Let us mention here the ideas around anomaly matching, used for exploration of model building, as well as the c-theorem. 

Some theories display symmetries that may be broken by an anomaly in the algebra of its operators, yet these anomalous terms are protected against renormalization. In this situation, the UV and IR realizations of the anomaly are strictly related, the anomaly central extension is identical. This feature serves the purpose of exploring possible UV completions of theories from their known IR realization. This scenario is different
in spirit to the $c-d$ conjecture, as the latter is not related to the possible presence of some symmetry algebra.

The $c$-theorem states 
the irreversibility of renormalization group  flows. In one dimension, it was originally proven that the central charge of an 
ultraviolet fixed point $c_{UV}$ is larger than the corresponding infrared one $c_{IR}$, that is $c_{IR}\le c_{UV}$. 
This theorem is rooted in unitarity \cite{Zam86,CFL91}. The extension of this theorem to higher dimensions is based on 
the conformal anomaly terms related to the stress tensor \cite{Ca88,FL98, KS11}. As mentioned before, the $c-d$ conjecture acts as a sort of $c$-theorem acting from a lattice representation to a conformal field theory one.

Let us elaborate on a particular case of a renormalization group  flow. Consider a local interaction such 
that some local high energy levels are not populated, that is, some local degrees of freedom decouple and play no role in the 
long-distance description of correlations. Then, it is possible to integrate these modes out, by just dropping them, 
and produce an effective description of the model in terms of the rest of modes. 
This description remains faithful at long distances. As a matter of fact the initial local dimension 
$d_{UV}$ is larger than the final one $d_{IR}$. Then along this RG flow
\beq
  d_{UV}\ge d_{IR} \Longrightarrow c_{\text{max,UV}} \geq c_{\text{max,IR}} \, .
\eeq
This idea is aligned with the $c-d$ conjecture.

Our conjecture has been applied to Hamiltonians with periodic boundary conditions, 
and we anticipate that a similar statement would hold for open boundary conditions. 
An intriguing question concerns the behavior under non-unitary Hamiltonians, 
where the central charge is replaced by the effective central charge. 
Additionally, it would be valuable to explore whether a similar conjecture extends 
to the chiral central charge of gapped 2D systems, following the approach in reference \cite{Kim}. 
Another promising research direction involves anyonic chains, where the Hilbert space lacks 
a tensor product structure and is instead described by a fusion tree \cite{Gils}. 
Lastly, we highlight an interesting result that proposes a similar lower bound on the central charge \cite{Bar}, 
though it depends on Lieb-Schulz-Mattis constraints with anomalies as a key factor—a condition that our general approach does not assume.
Some of these topics will be addressed in the future.

\vspace{0.2cm} 

\subsection{ Acknowledgements} 

We would like to thanks F. Alcaraz, M. Barkeshli, P. Calabrese, D. Bernard, G. Mussardo, B. Shi,  H.-H. Tu,  F. Verstraete, and J. Vidal  for conversations. GS 
acknowledges financial support through the Spanish MINECO grant PID2021127726NB-I00, 
the Centro de Excelencia Severo Ochoa Program SEV-2016-0597, 
the CSIC Research Platform on Quantum Technologies PTI-001,  
the QUANTUM ENIA project QUANTUM SPAIN, 
and the EU through the RTRP-Next Generation within the framework of the Digital Spain 2025 Agenda.


\section{SUPPLEMENTARY MATERIAL}

\subsection{Evidence for the $c-d$ conjecture}

We present below a collection of Hamiltonians of the form \eqref{3}
that are critical and described by a unitary CFT. The values of the local dimension of the Hilbert space, $d$,
the central  charge $c$, the corresponding CFT and symmetry  are collected in Table 1 in the main text. 

\begin{itemize} 

\item Ising Model \cite{BMD70,P70}
\beq
H_{\rm Ising} = - \sum_{i=1}^L  ( \sigma^x_i  \sigma^x_{i+1}  + \sigma^z_i)  \, , 
\label{m1}
\eeq 
where $\sigma^{x,z}_i$ are Pauli matrices. 
The associated CFT is the minimal model
${\cal M}_{3,4}$ \cite{BPZ84} with central charge $c= \frac{1}{2}$.

\item $Q$-state Potts model \cite{W82,B82}
\beq
H_Q = - \sum_{i=1}^L \sum_{n=1}^{Q-1} \left( ( X_i X^\dagger_{i+1})^n + Z^n_i \right)  \, , 
\label{m2}
\eeq
where $X$ and $Z$ are $Q \times Q$ matrices satisfying the relations 
$X^Q = Z^Q = {\bf 1}$, $X Z = e^{ 2 \pi i/Q} Z X$. The case $Q=2$
coincides with the Ising model. This model is critical for $Q=2,3,4$
with central charge $c_m = 1- \frac{6}{m (m+1)}$, where $\sqrt{Q} = 2 \cos( \frac{ \pi}{m+1})$.
The cases $Q=3$  and $Q=4$ correspond to $c_{5} = \frac{4}{5}$ \cite{D84}  and $c_{\infty}=1$ \cite{GR86}. 

\item XXZ$_{1/2}$  model \cite{B82} 
\beq
H_{XXZ} =  \sum_{i=1}^L \left( \sigma^x_i \sigma^x_{i+1}  +  \sigma^y_i \sigma^y_{i+1}  + \Delta
 \sigma^z_i \sigma^z_{i+1}  \right) \, , 
 \label{m7}
 \eeq
 where $\sigma^a_i$ are the Pauli matrices. 
 For any $\Delta$ there is a $U(1)$ symmetry 
 $\sigma^z \rightarrow \sigma^z$, $\sigma^\pm  \rightarrow e^{ \pm i \theta} \sigma^\pm$
 that is enlarged to $SU(2)$ at $\Delta =1$. 
  This model is conformal invariant  for  $-1 < \Delta \leq 1$ with $c=1$. It is described
  by a compact boson with radius $R = \sqrt{ \frac{2}{\pi} \arccos(- \Delta)}$ \cite{KB79,LP75}. 
  For $\Delta=0$ the Hamiltonian \eqref{m7}  can be mapped, via a Jordan-Wigner transformation, into a hopping Hamiltonian of a spinless fermion.    At $\Delta=1$,
  the CFT is described by the WZW model $SU(2)_k$  at level $k=1$ \cite{AA86}.

\item Ashkin-Teller model \cite{AT43}
\barray 
H_{AT} &  =  &  - \sum_{i=1}^L \left( \sigma^z_i + \tau^z_i + 
\Delta \sigma^z_i \tau^z_i \right.   \label{m3}  \\ 
& + & \left. \sigma^x_i \sigma^x_{i+1} + \tau^x_i \tau^x_{i+1} + 
\Delta  \sigma^x_i \sigma^x_{i+1} \tau^x_i \tau^x_{i+1} \right) \, ,  
  \nonumber 
\earray 
where $\sigma^a_i$ and $\tau^a_i$ are two sets of commuting Pauli matrices. 
This model describes the coupling of two critical Ising models via the four spin term. 
It has $\mathbb{Z}_2 \times \mathbb{Z}_2$ symmetry. The model is critical in the line
$-1/\sqrt{2} < \Delta \leq 1$ corresponding to a $Z_2$-orbifold boson with radius $R_0 = \sqrt{ \pi/( 2 \arccos(- \Delta)})$ 
\cite{KNK81,Y87,S87,Bv87}


\item Parafermionic $Z_Q$ \cite{FZ82} 
\beq
H_Q = - \sum_{i=1}^L \sum_{n=1}^{Q-1} \frac{1}{ \sin( \pi n/Q)}  \left( ( X_i X^\dagger_{i+1})^n + Z^n_i \right) 
\label{m4}
\eeq
where $X$ and $Z$ denote identical matrices as those used in the $Q$-state Potts model.  
This model is conformal invariant for any $Q\geq 2$ with $c_Q = 2 (Q-1)/(Q+2)$. 
For $Q=2$ and 3 one recovers the Ising and the 3-state Potts models, while
for $Q=4$ one finds the Ashkin-Teller model with $\Delta = 1/\sqrt{2}$ \cite{AL86,A87}.

\item Fateev-Zamolodchikov model  \cite{FZ80} 
\barray 
H_{FZ} & = &  \varepsilon \sum_{i=1}^L 
\left[  \sigma_i - (\sigma^z_i)^2 - 2 ( \cos \gamma -1) ( \sigma^\perp_i  \sigma_i^z + \sigma^z_i  \sigma_i^\perp ) \right.  \nonumber  \\
& - & \left.   2 \sin^\gamma ( \sigma^z_i - (\sigma^z_i)^2 + 2 (S^z_i)^2 
\right] \label{m5} \, , 
\earray 
where $\vec{S} = (S^x, S^y, S^z)$ are the spin 1  $SU(2)$ matrices,
$\sigma^z_i = S^z_i S^z_{i+1}$ and  $\sigma_i = \vec{S}_i \cdot \vec{S}_{i+1} = \sigma^z_i + \sigma^\perp_i$.
The model is critical for $0 \leq \gamma \leq \frac{\pi}{2}$, with $c= \frac{3}{2}$ for $\varepsilon = +1$ \cite{AM88a,FSZ88}
(antiferromagnetic model) and $c = 1$ for $\varepsilon = -1$  \cite{AM89}(ferromagnetic model) . 
  
\item    Higher spin XXX$_{s}$ model \cite{higherS}
\beq
H_{XXX_s} = \sum_{i=1}^L \sum_{k=1}^{2s} 
a_k P_k( \vec{S}_i +  \vec{S}_{i+1}) \, , 
\eeq
where $P_k$ are projectors out spin $k$ and 
the constants are given by 
\beq 
 a_k  =  \sum_{\ell=1}^k \frac{1}{\ell} \, . 
 \nonumber 
 \eeq
The case $s=1/2$ coincides with the Heisenberg Hamiltonian $\Delta=1$  (\ref{m5}).
The spin 1 and 3/2 Hamiltonians  are given, up to an  additive constant, by 
\barray 
H_{XXX_{1}} & = & \frac{1}{4}  \sum_{i=1}^L \left[  \vec{S}_i \cdot \vec{S}_{i+1}  - (  \vec{S}_i \cdot \vec{S}_{i+1} )^2 \right] , 
\nonumber \\
H_{XXX_{3/2}} & = & \sum_{i=1}^L \left[  - \frac{1}{16} 
 \vec{S}_i \cdot \vec{S}_{i+1}  + \frac{1}{54}  (  \vec{S}_i \cdot \vec{S}_{i+1} )^2   \right. \nonumber \\  
& +  &   \left.  \frac{1}{27} 
 (  \vec{S}_i \cdot \vec{S}_{i+1} )^3  \right] . \nonumber 
\earray 
The underlying CFT is given by the WZW model $SU(2)_{k}$ with $k=2s$ \cite{AA86,AH87,AG88}

\item Spin 1 bilinear-biquadratic Heisenberg model
\beq
H = \sum_{i=1}^N  \left[ 
\cos \theta \; ({\bf S}_i \cdot  {\bf S}_{i+1} )
+ \sin \theta \; ({\bf S}_i\cdot{\bf S}_{i+1})^2
\right]  \, , 
\eeq
where ${\bf S}_i$ are spin 1 matrices. 
As a function of the angle $\theta$, this model exhibits several critical phases.  At $\theta = -\frac{\pi}{4}$ it coincides with the spin 1 XXX model considered above that has $c = \frac{3}{2}$, while at $\theta = \frac{\pi}{4}$ it is described by the $SU(3)_1$ model with $c=2$ \cite{IK97}. The entire region $\lambda \in \left[ \frac{\pi}{4}, \frac{\pi}{2} \right)$ is critical with $c = 2$. Further extensions to $SO(N)$ can be found in \cite{R85,TO11}.

\item $SU(N)$ model \cite{S75} 
\beq 
H = \sum_{i=1}^L P_{i, i+1} \, , 
\eeq
where $P_{i,i+1}$ is the permutation operator acting on $N$ values of spins (colors) that corresponds to the fundamental representation of $SU(N)$. 
For $N=2$ this Hamiltonian reduces to Heisenberg spin $1/2$. 
The low energy physics of this model is described by the WZW model $SU(N)_1$ which
has central charge $c= N-1$ \cite{A88,BS99,A99}. 

\item{Blume-Capel model  \cite{A85,XA11}}

\beq
H_{BC} = - \sum_{i=1}^L \left( S^z_{i} S^z_{i+1} - \delta (S^z_i)^2 - \gamma S^x_i \right)
\label{m6} 
\eeq
where $S^x, S^z$ are spin-1 $SU(2)$ matrices.  For $\gamma > \gamma_{\rm tr} \simeq 0.41653$
the model has a critical line $\delta_c(\gamma)$ in the Ising universality class, $c=1/2$. 
At $\delta_{\rm tr} \simeq 0.91024$ there is a tricritical point described by the tricritical
Ising model with $c= 7/10$.



\end{itemize}

\end{document}